\documentclass{iopart}
\usepackage[]{graphicx}
\begin{document}

\title{Mesoscopic fluctuations of spin currents}
\author{Yuli V. Nazarov}
\address{Kavli Institute of NanoScience, 
Delft University of Technology, \\
2628 CJ Delft, the Netherlands}
\begin{abstract}
Spin currents may be generated by applying bias
voltages $V$ to the nanostructures even in the absence 
of spin-active ferromagnetic interfaces. Most theoretical
proposals concentrate on a concrete spin-orbit
interaction and on the disorder-averaged effect.
It remains underappreciated that any spin-orbit
interaction produces random spin currents with 
a typical amplitude $\simeq e^2 V/\hbar$ not affected
by disorder. 
This work addresses such mesoscopic fluctuations
of spin currents for generic model of a nanostructure
where several quantum connectors meet in a single node.
The analysis is performed in the framework of
recently developed quantum circuit theory of $G_Q$ corrections
and reveals four distinct mechanisms of spin current fluctuations.
The results are elaborated for  simple models of 
tunnel and ballistic connectors.
\end{abstract}
\pacs{0000}
\submitto{\NJP}
\maketitle
\section{Introduction}
In recent years considerable theoretical and experimental 
work is aimed at the controlled manipulation of electron spin 
in (nanoscale) solid state systems, a 
field commonly referred to as \textit{spintronics}~\cite{dassarma}. 
Spin current defined as a flow of electron spin is one
of the central and most useful concepts of this field.\cite{bauerreview}
Much confusion can be easily brought by irresponsible 
and/or uncareful use of this concept. This confusion is
related to two facts. Firstly, 
spin currents, unlike electric or particle
currents, are not usually conserved under 
interesting circumstances. Secondly, it is frequently
not evident that the spin currents due to electrons
near Fermi surface really present
a dominant contribution in spin balance: This is 
certainly not true for equilibrium and/or electrically isolated
systems.

Therefore, it seems prudent to start the
article with a short description of the measurement 
that can directly access such spin current \cite{bart} and/or 
its temporal fluctuations\cite{Antonio1}
Let us consider a current injector of the resistance $R$ that
pushes spin-polarized electrons
into a metallic lead. In principle, this polarization disappears
far in the lead. However, spin relaxation is usually a slow
process so the spin current  is preserved at significant
distance $L_{sf}$ from the injector. Let us set up a 
{\it spin-active} interface within this distance from the
injector. A spin-active interface provides 
different resistance for electrons of two different spin projection
on the polarization axis of the interface. 
Most common example is the interface between
a ferromagnetic and non-ferromagnetic metal. In this case, the 
polarization axis coincides with the direction of magnetization 
of the ferromagnet. 

The voltage drop over the interface is thereby
sensitive to spin of the electrons that traverse it.
We would like the
difference of the voltage drops at {\it opposite} directions
of the polarization axis is directly proportional
to the projection of the spin current on the polarization axis.
This enables the measurement of spin current as a voltage measurement.
To achieve this, one has to choose the interface resistance $R_{in}$
such that $R_{in} \ll R$ but still by far exceeds the
resistance of the metal piece between the 
injector and the interface. We shall stress the useful analogy between
spin current and spin accumulation, on one side, and electric
current and voltage on another side. The above condition on $R_{in}$
is precisely the same as the condition
for the measuring the electric current
from the injector through the voltage drop over the interface.

The disadvantage of this measurement scheme is that at a given
polarization of the interface only a single projection
of the spin current can be measured. The point is 
that the spin current
with the projection perpendicular to the polarization axis
is usually immediately adsorbed at the interface.\cite{bauerreview}
An alternative scheme \cite{Antonio2}
where all three components of the spin current can be measured
in the least obtrusive way is still difficult to realize 
for solid-state devices.

Spin currents in the systems where spin-active 
interfaces dominate the resistance, are well-studied. \cite{bauerreview}
Usually, a simple circuit-theory approach based on the
balance of spin and charge currents \cite{spin-circuit}
suffices to describe everything.
Here, we consider the situation where there are no spin-active
interfaces in the nanostructure used as an injector of the
spin current. Still there is spin-orbit interaction.

It was noticed quite long ago \cite{eliashberg} that in bulk homogeneous
materials without inversion center electric currents produce
spin currents. However, in bulk materials this effect is usually
negligible in comparison with other agents of spin orientation,
for instance, with the effect of magnetic field produced by
the currents.\cite{surface} In ballistic nanostructures,
the spin-orbit interaction in combination with geometric
effects may lead to spin currents. Consider, for instance,
a single isotropic scattering center subject to electron flow
in $z$ direction. Intensity of electron scattering to $+x$ and $-x$
directions will depend on the spin projection on $y$ axis.
If we direct the scattered electrons to different terminals,
we get spin currents in the terminals. Such device proposal
has been actually elaborated.\cite{kiselev}
At more abstract microscopic level, spin-orbit interaction
always results in a spin-dependent scattering matrix (see, for instance,
\cite{pareek}). However, since the effect relies on geometry,
it is drastically decreased by any scattering in the nanostructure.
Indeed, the scattering makes electron distribution function isotropic
so that they forget the geometry. In scattering matrix
language, the spin currents are determined by
random phase shifts of the scattering amplitudes. After averaging over random phase shifts, the average spin current
is supposed to average out if the scattering is sufficiently
chaotic. This has been confirmed
by extensive numerical experiments.\cite{kramer1}

It remains underappreciated that even for
completely chaotic scattering there must be significant
mesoscopic fluctuations of spin currents. They are 
governed by the same quantum interference mechanism as
the celebrated Universal Conductance Fluctuations \cite{Altshuler}
and must be of the same scale. Such fluctuations are clearly
seen in numerical simulations at quantum level \cite{kramer2}.
Recent numerical work \cite{Ren} provides a detailed study
of these fluctuations for a concrete device.

In this work, we theoretically investigate the mesoscopic fluctuations
of spin currents for multi-terminal nanostructures assuming
sufficient isotropization of electron distribution function
that allows to disregard all geometric effects. This means
that after averaging over random phase shifts/impurity positions
there is no net spin effect and the system is $SU(2)$ invariant
with respect to rotations in spin space. Under
these conditions, spin-orbit
interaction only manifest itself in spin relaxation and completely
characterized by the rate of such relaxation.
We denote the full spin current with  spin index $\alpha=1,2,3$ coming
to the terminal $i$ as $I^{\alpha}_i$.
We study the correlators of the two,
$$
\langle\!\langle I^{\alpha}_i I^{\beta}_j\rangle\!\rangle = 
\delta_{\alpha\beta}I_{ij} 
$$
that are $SU(2)$ invariant. 

We stress that we study spin current fluctuations
that persist at least at the scale of the contact size. This
is very different from mesoscopic fluctuations of
spin accumulation. Such fluctuations are known
to form a random pattern correlated at the scale of
mean free path \cite{nazarov-polarization,zuyzin} and
are disregarded here.
\section{Model, parameters, expectations}
We model a nanostructure along the lines 
of quantum circuit theory \cite{circuittheory,Gabriele}
representing a complicated setup in terms of
three kinds of the elements: terminals, nodes,
and connectors. An arbitrary nanostructure with sufficiently
chaotic scattering inside can be presented in this way
with using sufficiently many nodes and connectors. 
\begin{figure}
\centerline{\includegraphics[width=12cm]{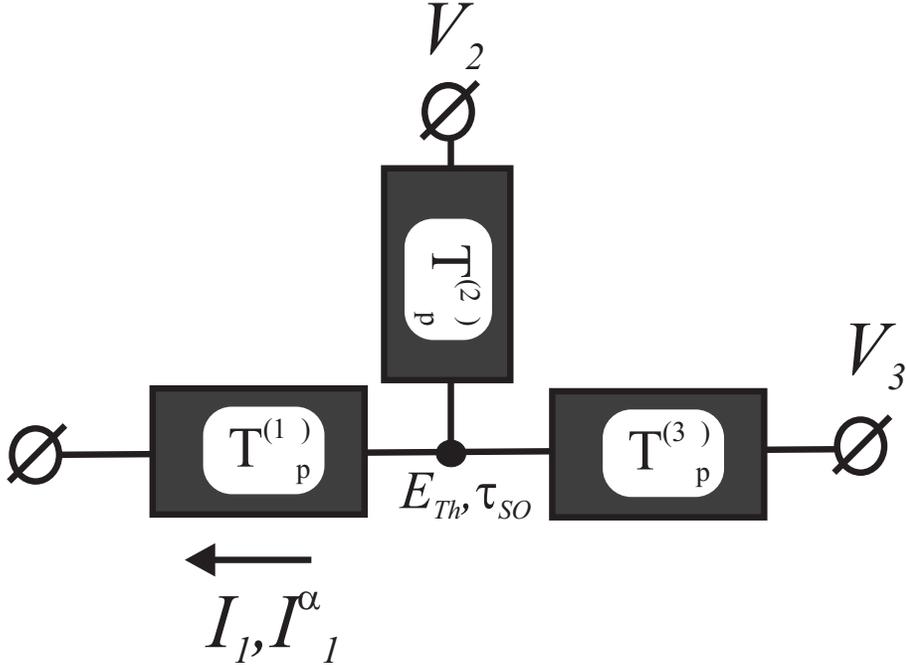}}
\caption{Generic model of nanostructure. We consider
a multi-terminal setup (three terminals shown in the Figure).
The connectors are characterized by the sets of transmission
eigenvalues ($T^{(i)}_p$). The node is characterized by
Thouless energy related to the electron dwell time in the
node. The spin-orbit interaction is represented 
by $\tau_{SO}$, spin relaxation time in the node.}
\label{setup}
\end{figure}
The specific model under consideration 
consists of a single node attached to many reservoirs 
by arbitrary connectors (Fig. \ref{setup}). The node
in circuit theory is very close to "chaotic cavity"
used in Random Matrix Theory (RMT) of quantum transport.
\cite{BeenakkerReview}

The connector $i$ is characterized by
a set of transmission eigenvalues $T^{(i)}_p$, $p$ labelling
quantum channels in the connector.
Conductance of the connector reads 
$G^{(i)}=G_Q\sum_p T^{(i)}_p$. In the limits of applicability
of the circuit theory, $G^{(i)}_T\gg G_Q$.
There is a single terminal per connector most generally
characterized by an energy-dependent electron filling
factor $f_i(E)$.
The mesoscopic fluctuations of two-terminal conductance in this 
model have been considered
in \cite{Brouwer} by RMT methods.

Spin current fluctuations can not be accessed by RMT
since, as we will see, they disappear in the limit of
pure ensembles. We need more parameters that are
absent in RMT but present in more detailed and microscopic
theories \cite{Altshuler} that describe transitions between
the ensembles. 
The size of the node is characterized by the mean level
spacing $\delta_S$, or, more adequately, 
by the Thouless energy
$E_{th} \equiv G_\Sigma \delta_S/G_Q$, $G_\Sigma \equiv =\sum_i G^{(i)}_T$.
Mesoscopic fluctuations are supposed to correlate
at energies $\simeq E_{th}$.
For the present purposes, the most 
important parameter is the spin-orbit 
time $\tau_{SO}$ that determines spin relaxation in the node.
We assume no spin-orbit effects in the connectors.
Therefore, $\tau_{SO}$ contains all information
concerning the spin-orbit interaction and its strength is given
by a dimensionless $\eta_{SO} \equiv \hbar/\tau_{SO}E_{th}$.

At moderate values of this parameter, $\eta_{SO} \simeq 1$,
we expect the mesoscopic fluctuations of spin current
to be very similar to the mesoscopic fluctuations
of the charge current. Those are known
to be universal corresponding to the fluctuating conductance
of the order of $G_Q$. At low voltages $eV \ll E_{th}$ 
this gives $\langle\!\langle I^2\rangle\!\rangle \simeq G^2_Q V^2$
while at large voltages $eV \gg E_{th}$ 
$\langle\!\langle I^2\rangle\!\rangle \simeq G^2_Q V E_{th}/e$
\cite{LarkinKhmelnitskii}.
The distinction is that the spin current 
is not conserved. Therefore, we expect non-zero fluctuations
of the total spin current $I_{\alpha} =\sum_i I^{(i)}_\alpha$
to/from the nanostructure. Those must be 
of similar magnitude as the fluctuations of the current to/from
a reservoir.

Naturally, we expect the spin current fluctuations
to vanish in the limit $\eta_{SO} \ll 1$ 
of weak spin-orbit interaction. The opposite limit
is less obvious: One might think that strong spin-orbit
interaction leads to fast spin relaxation in the node,
the latter obviously suppresses any spin effects. 

The results derived and discussed in the article
conform to these general expectations and give details
of spin current fluctuations and their correlations
for connectors of low and high transmission.

\section{Circuit-theory action}
Qualitatively, the model is most conveniently described
by the action ${\cal S}$ that depends on the 
matrix Green's function of the node $\check{G}$
and those of the reservoirs, $\check{G}^{(i)}$.
All such matrices obey the constrain $\check{G}^2=\check{1}$.
It is also known that the concrete matrix structure
of the matrices does not have to be specified at this stage:
most terms in the action do not depend on concrete physical
realisation of such structure. \cite{Gabriele}

The action reads:
\begin{eqnarray}
{\cal S}= {\cal S}_J + {\cal S}_E + {\cal S}_{so};\label{model}\\
S_J^{(i)} = \frac{1}{2} \sum\limits_p 
{\rm Tr}\left[ \ln\left( 1+ \frac{T_p^{(i)}}{4}(
\check{G}^{(i)}\check{G} +\check{G}\check{G}^{(i)} -2)\right)\right] 
 \\
{\cal S}_{E} = -\frac{i\pi}{\delta_S}{\rm Tr}[\check{E}\check{G}]\\
{\cal S}_{so} = \frac{\pi}{8\delta_S\tau_{so}} 
{\rm Tr}[\check{G}\check{\sigma}_{\alpha}\check{G}\check{\sigma}_{\alpha}  ]
\end{eqnarray}
Here, ${\cal S}_i$ gives a contribution of an individual
connector that depends on the matrix Green's functions
$\check{G}$
, $\check{G}^{(i)}$ at
its ends and the transmissions of the transport channels.
Next term, ${\cal S}_{E}$, accounts for the finite size
of the node and therefore for the de-correlation of 
the mesoscopic fluctuations $\langle\!\langle A(E)A(E')
\rangle\!\rangle$
upon increasing energy difference $E-E'$.
The last term presents  spin-orbit
effects, $\sigma_\alpha$ being Pauli matrices in spin space.

We make use of Keldysh technique, our action approach is
similar to \cite{LarkinKeldysh}
The Green functions in the terminals at given energy $E$
have the standard Keldysh structure
\begin{equation}
\check{G}_i=\left(\begin{array}{ll}
1-2 f_i & 2f_i \cr
2-2f_i  & -1+2f_i
\end{array} \right)
\label{Keldysh-Green}
\end{equation}
$f_i(E)$ being the energy-dependent filling factor
in each reservoir, and are diagonal in spin index.
The optimal point of the action gives the 
equilibrated Green function $\check{G}_N$ in the node.
It has the same structure as (\ref{Keldysh-Green})
with the {\it balanced}
filling factor $f_N = \sum_i G^{(i)}_T f_i /\sum_i G^{(i)}_T$.
From now on, it is convenient for us not to distribute
explicitly transport channels over the connectors. We achieve
this by labelling all transport channels with $i$ and
ascribing a terminal with filling factor $f_i$ to each transport
channel. In these notations, $f_N =\sum_i T_i f_i /\sum_i T_i$.

To account for spin currents to/from each reservoir, we perform 
rotations with spin-dependent counting field $\chi^\alpha_i$
in each terminal\cite{rotations},
\begin{equation}
\check{G_i}(\chi^\alpha_i) = 
\exp\left(i\chi^\alpha\sigma^\alpha\tau_z\right/2)
\check{G_i}
\exp\left(-i\chi^\alpha\sigma^\alpha\tau_z\right/2)
\end{equation}
In this case, the action will have a non-trivial 
saddle-point $\check{G}_N(\{ \chi^{\alpha}_i\}$
with the $\chi$-dependent optimal value of the action. 
The expansion in $\chi_\alpha$ gives the momenta of
low-frequency temporal fluctuations of spin currents.\cite{Antonio2} 

However, such temporal fluctuations is not 
a problem of present interest.
Firstly, we are interested
in the time-averaged spin current only, so we only have
to keep the first term in $\chi$-expansion of each Green's
function,
\begin{equation}
\check{G}_i(\chi^\alpha_i) \approx \check{G}_i + 
\delta\check{G}_i; \delta\check{G}_i = 2\chi_i^\alpha\sigma^\alpha 
\left(\begin{array}{ll}
0 & f_i \cr
f_i-1  & 0
\end{array} \right)
\end{equation}

Secondly, this time-averaged current vanishes in the 
main order in $G/G_Q$ so we shall turn to the analysis
of so-called $G_Q$-corrections corresponding to mesoscopic
fluctuations. The general setup for the evaluation of
$G_Q$ corrections in the framework of circuit theory 
has been put forward in Ref.\cite{Gabriele}
and, as it is accustom for fluctuations, involves quadratic
expansion of the action in the vicinity of the saddle point.
It goes as follows. We wish to access the correlation
of the mesoscopic fluctuations at two 
different values of a parameter set: we call these values 
"black" and "white". For us, the parameter set in principle
includes energy $E$, filling factors $f_i$ at this energy
and counting fields $\chi^\alpha_i$. We find 
the saddle points $G^w_N,G^b_N$ corresponding for each set of 
parameters ( $E_w,f^w_i,\chi^{w,\alpha}_i$ and $E_b,f^b_i,
\chi^{b,\alpha}_i$). We {\it double} the matrix structure
so that the node Green function consists of four
blocks: two diagonal $ww$,$bb$ and two non-diagonal $wb$,$bw$.
We give $\check{G}$ a fluctuation $\check g$
concentrated in the non-diagonal blocks. The constrain $\check{G}^2=1$
is satisfied up to the second order in $\check{g}$ if
we substitute 
$$
\check{G} =\check{G}_N + \check{g} - \check{g}^2\check{G}_N/2
$$
under the constrain 
\begin{equation}
\check{g}\check{G}_N +\check{G}_N \check{g} =0.
\label{g-constrain}
\end{equation}
If we introduce a compound index $\bar{a}$ composed of two check
indices, the result of quadratic expansion can be written as
\begin{equation}
\delta {\cal S} = g^{wb}_{\bar{a}} M_{\bar{a}\bar{b}} g^{bw}_{\bar{b}} - 
\eta_{\bar{a}}g^{bw}_{\bar{a}}.   
\label{M-action}
\end{equation}
To take proper care of the constrain (\ref{g-constrain}),
we  consider matrix $\Pi_{\bar{a}\bar{b}}$
defined trough the following relation:
\begin{eqnarray}
\Pi_{\bar{a}\bar{b}} g^{bw}_{\bar{b}}\rightarrow \nonumber\\
\frac{1}{2}\left(\check{g} - \check{G}_0 \check{g} \check{G}_0\right)\rightarrow \nonumber\\
\frac{1}{2}\left(\check{g}^{bw} - \check{G}_b \check{g}^{bw} \check{G}_w \right),
\end{eqnarray}
the last equation makes white-black block separation explicit. We note
that $\Pi_{\bar{a}\bar{b}}$ is a {\it projector}: It separates "bar" space
on two subspaces where $\check{g}$ either 
commutes or anti-commutes with $\check{G}_N$,
and projects an arbitrary $\check{g}$ onto anti-commuting subspace.
The matrices $M_{\bar{a}\bar{b}},\Pi_{\bar{a}\bar{b}}$
define the fluctuation correction ${\cal S}_{G_Q}$ sought:
\begin{eqnarray}
{\cal S}_{G_Q}  = \ln {\rm det} 
\left( \tilde{M}_{\bar{a}\bar{b}}\right);\\ 
 \tilde{M}_{\bar{a}\bar{b}} =\Pi_{\bar{a}\bar{b}} M_{\bar{b}\bar{c}}
 \Pi_{\bar{c}\bar{b}} +\delta_{\bar{a}\bar{b}}-\Pi_{\bar{a}\bar{b}}.
\label{in-M-and-P}
\end{eqnarray}

The quantity of our interest is the correlator
of two spin currents per energy interval. Since a single
current is given by a term linear in $\chi$,
the correlator is given by the term which is linear in both
$\chi^w$ and $\chi^b$,
\begin{eqnarray}
\langle\!\langle j^{\alpha}_i(E_b) 
j^{\beta}_k(E_w)\rangle\!\rangle = 
\mathop{\lim\limits_{\chi^{w,b}\to 0}}
\frac{\partial{\cal S}_{G_Q}}{\partial\chi^{b,\alpha}_i.
\partial\chi^{w,\beta}_k} 
\label{main-limit}
\end{eqnarray}

Eventually, this gives diffuson contribution to $S_{G_Q}$.
The Cooperon contribution is obtained in similar manner
with one of the Green's functions ($\check{G}_w$ or $\check{G}_b$)
transposed. However, our particular model is invariant
with respect to time reversal. Diffuson and Cooperon contributions
to spin current fluctuations are therefore identical and it is
enough to calculate only one of the two.


\section{Strategy and workout}
The calculations by the method described 
are straightforward but, admittedly, rather lengthy.
The correct choice of strategy is vital for speedy
evaluation.  An obvious strategy
is to follow \cite{Gabriele} in detail: compute
the fluctuation action by expanding near the 
$\chi$-dependent saddle point, get it as a function
of $\chi$, go to the limit (\ref{main-limit}).
However, such straightforward approach 
An alternative strategy is to expand the Green function near the 
"trivial" saddle-point solution $\chi=0$. Since
the $\chi$-dependent saddle point is close to trivial point,
it will be sooner or later covered by such expansion.
In fact, this approach is
very close to the traditional Cooperon-Diffuson technique.
\cite{LarkinKhmelnitskii} A disadvantage is that in this case 
one has to expand
to at least third order in $\check{g}$ and satisfying the
constrain in this order is a headache.

Here, we adopt a mixed strategy. We expand in the $\chi$-dependent
point but keep $\chi$ small. This amounts to expanding
the matrix $\tilde{M}$ from Eq. \ref{in-M-and-P} to the orders in $\chi^{w,b}$ we need,
\begin{equation}
\tilde{M}=\tilde{M}_0 + \tilde{M}_w + \tilde{M}_b +\tilde{M}_{wb},  
\end{equation}
where
$$
\tilde{M}_{w,b} \propto \chi^{w,b};\; \tilde{M}_{wb} \propto \chi^{w}\chi{w},  
$$
Expanding the log in Eq. \ref{in-M-and-P}
we observe that the term we need is given by
\begin{equation}
{\cal S}_{G_Q} = {\rm \bar{Tr}}\left[ \tilde{M}_{wb}
(\tilde{M}_0)^{-1} - \tilde{M}_w (\tilde{M}_0)^{-1} \tilde{M}_b(\tilde{M}_0)^{-1}\right]
\label{SGQ-expanded}
\end{equation}
We also observe that elements of the matrix $(\tilde{M}_0)^{-1}$
correspond to "common" diffuson propagators of the traditional
mesoscopic fluctuation technique. Two terms in (\ref{SGQ-expanded})
contain one and two diffuson propagators and in traditional
theory describe respectively local and distant current correlations.

An essential element of the strategy is to evaluate $\tilde{M}$
in a convenient basis. We choose the basis in Keldysh-spin space
in such a way that $G_N(\chi)$ always remains diagonal.
At $\chi=0$ this is achieved  by the transformation
\begin{equation}
\check{P}^{-1} \check{G_N} \check{P}=-\tau_z; \; \check{P} = 
\left(\begin{array}{ll}
f_N & 1 \cr
f_N-1  & 1
\end{array} \right).
\end{equation}
separately in black and white  sector. The basis rotates upon changing
$\chi^{w,b}$, but this rotation is easy to take into account 
perturbatively. This basis is especially convenient
to satisfy the anticommutation constrain 
(\ref{g-constrain}): It is satisfied by 
\begin{equation}
\check{g} = 
\left(\begin{array}{ll}
0 & g_+ \cr
g_-  & 0
\end{array} \right).
\end{equation}
for both non-diagonal blocks.
Owing
to SU(2) symmetry in spin space, spin structure
of the mesoscopic fluctuation in a given block(say, bw)
can be most conveniently
presented in terms of the singlet and
three triplet components,$g^{bw}+ g^{bw}_\alpha\sigma_\alpha$.
The matrix $\tilde{M}_0$  does not 
contain elements that mix the components.
For singlet-singlet elements we obtain
\begin{eqnarray}
(\tilde{M}_0)_{+-}=((\tilde{M}_0)_{-+})* = \nonumber \\
\frac{1}{4}\left(\sum_i T_i\right) (1-i(E_b-E_w)/E_{th}); \nonumber \\
(\tilde{M}_0)_{++}= \frac{1}{4}\sum_i T_i^2 \tilde{f}^b_i\tilde{f}^w_i.
\end{eqnarray}
Here we introduce conveniently physical notations:
filling factor drops over the transport channel $i$,
$\tilde{f}_i \equiv f_i -f_N$. For instance, the charge
current per energy interval in this channel is given
by $(2\pi)^{-1}T_i\tilde{f}_i$. For triplet-triplet elements, $\eta_{SO}$
comes into play:
\begin{eqnarray}
(\tilde{M}_0)_{+-}=((\tilde{M}_0)_{-+})* = \nonumber \\
\frac{1}{4}\left(\sum_i T_i\right) (1-i(E_b-E_w)/E_{th});
\end{eqnarray}
Further calculations
give $\tilde{M}_{b,w},\tilde{M}_{wb}$. We give expressions
for this matrices skipping the trivial 
spin structure and spin index of $\chi_i$. We also skip $++$ elements
since latter do not contribute to the trace (\ref{SGQ-expanded}).
\begin{eqnarray}
\tilde{M}^{b}_{+-}=\tilde{M}^b_{-+} = \nonumber \\
\frac{T_i}{4} (1-T_i) \tilde{\chi}^b_i \tilde{f}^{b}_i
\end{eqnarray}
and similar for $\tilde{M}_w$. Here we simplify the 
answer
by introducing counting field "drops" over each connector
\begin{equation}
\tilde{\chi}_i = \chi_i -\chi_N, \chi_N = \frac{\sum_i T_i \chi_i}
{(\sum_i T_i)(1+ \eta_{SO})}
\label{chiN}
\end{equation}
The above relations are similar to those for filling factors
apart from the factor $(1+\eta_{SO})$ that accounts
for spin relaxation in the node. Without the factor,
the structure guarantees the current conservation: If $\chi_i$
are the same in each terminal, all the "drops" $\tilde{\chi}$ vanish.
As to $\tilde{M}_{wb}$,
\begin{eqnarray}
\tilde{M}^{wb}_{--} =\frac{1}{4}\sum_i T_i^2 \tilde{\chi}^b_i\tilde{\chi}^w_i
+\frac{\sum_i T_i}{32} \eta_{SO} \chi_N^b \chi_N^w ; \\
\tilde{M}^{wb}_{+-}=\tilde{M}^{wb}_{-+} =
\frac{1}{4}\sum_i T_i^2(1-T_i) \tilde{\chi}^b_i \tilde{f}^{b}_i\tilde{\chi}^w_i \tilde{f}^{w}_i.
\end{eqnarray}
Thereby we give the most general answer for the fluctuations
of the spin current
\begin{eqnarray}
\sum\limits_{i,k}\langle \!\langle j_w(E_w)j_b(E_b)\rangle\!\rangle \chi^w_k \chi^b_i = ({2\pi})^{-2}(\nonumber \\
A_bA_w {\rm Re}\left(D_t(D_s-D_t)\right) + \label{mech1}\\
A_{wb} {\rm Re} \left(D_s-D_t\right)+ \label{mech2} \\
+ F B_{wb} \left(|D_s|^2-|D_t|^2\right) \label{mech3} \\
+\frac{F}{8} \eta_{SO}\chi^w_N \chi^b_N \left(|D_s|^2-|D_t|^2\right) ) \label{mech4}
\end{eqnarray}
where 
\begin{eqnarray}
A_{b,w} = \sum_i T_i(1-T_i) \tilde{\chi}^{b,w}_i \tilde{f}^{b,w}_i/\sum_i T_i\nonumber\\
A_{wb} = \sum_i T_i^2(1-T_i) \tilde{\chi}^b_i \tilde{f}^{b}_i\tilde{\chi}^w_i \tilde{f}^{w}_i/\sum_i T_i\nonumber\\
B_{wb} = \sum_i T_i^2 \tilde{\chi}^b_i\tilde{\chi}^w_i/\sum_i T_i\nonumber\\
F= \sum_i T_i^2 \tilde{f}^b_i\tilde{f}^w_i/\sum_i T_i\nonumber\\
D_s=\frac{1}{1+i(E_b-E_w)/E_{Th}}; \nonumber\\
D_t= \frac{1}{1+i(E_b-E_w)/E_{Th}+\eta_{SO}}
\end{eqnarray}
It is instructive to confront this with the corresponding
expression for the fluctuations of the {\it charge} currents,
\begin{eqnarray}
\sum\limits_{i,k}\langle \!\langle j_w(E_w)j_b(E_b)\rangle\!\rangle \chi^w_k \chi^b_i = \nonumber (2\pi)^{-2}(\\
A_bA_w {\rm Re}\left(D^2_s +3 D^2_t)\right) + \label{mech1c}\\
A_{wb} {\rm Re} \left(D_s+3D_t\right)+ \label{mech2c} \\
+ F B_{wb} {\rm Re} \left(|D_s|^2+3|D_t|^2\right) \label{mech3c}
\end{eqnarray}
where the definition of $\tilde{\chi}$
in constants $A,B$ differs from (\ref{chiN}) by setting $\eta_{SO}=0$.  

The correlators are linear in both $f^w,f^b$ as it should be since
each (spin) current is linear in filling factors.

\section{Discussion of the general result}
We see that the spin current fluctuations are 
in fact contributed by three distinct mechanisms
corresponding to three terms (\ref{mech1}),(\ref{mech2}) and (\ref{mech3}).

First mechanism is the only one that provides fluctuations
for the system of tunnel connectors where all $T_i \ll 1$.
Inspecting its form, we find that the fluctuating
spin currents in each transport channels 
are always proportional to the corresponding filling factor
drops, and, apart from $(1-T_i)$ factor, to the corresponding
charge currents. So it looks like  the fluctuating
quantity is the polarization: The ratio of conductances 
of the channel, and not the spin/charge arriving
to the transport channel.
Moreover, the form of the term suggests that these polarization
fluctuations are "globally" orchestrated: The polarization
fluctuation is the same for all transport channels.  
Let us call the corresponding mechanism {\it global polarization
fluctuations}(GPF). Such "globality" is in agreement 
with the fact that in a common diffuson technique such
two-diffuson diagramms describe mesoscopic correlations at large
distances. We also see the same global correlation in corresponding
contribution to the charge current fluctuations: in this case,
it looks like the conductances of all transport channels 
fluctuate all together by the same value. Since fast 
spin relaxation in the node destroys the "global" coordination,
the mechanism ceases to work in the limit of large $\eta_{SO}$.

The polarization fluctuations do not have to
be correlated at global level: One can easy imagine
that they arise locally and do not correlate in different
transport channels.
This is how the second mechanism (\ref{mech2}) works.
We see that also in this case the current fluctuation in
each channel is proportional to the corresponding 
filling factor drops. This proves that the fluctuation is
a change of the polarization. However, we also see that in
contrast to (\ref{mech1}),
(\ref{mech2}) is contributed  by the products of counting
field drops at the same channel only, 
$\tilde{\chi}_i\tilde{\chi}_i$. 
Therefore, they do not 
correlate in different channels.
Let us call it {\it local polarization fluctuations}(LPF).
Due to the locality, the spin current fluctuations survive
in the limit of large $\eta_{SO}$.
Similar mechanism (Eq. \ref{mech2c}) provides uncorrelated fluctuations
of channel conductance for charge currents. As to $T_i$ dependence
of the mechanism, the origin of $(1-T_i)$ factor is clear: The
channel with $T_i=1$ has the maximum possible conductance
for both spin directions, so the 
fluctuations of conductance/polarization
must be suppressed. The origin of $T^2_i$ factor is not clear for the
author. However, the factor is here and owing to this 
the LPF mechanism does not work in the limit of tunnel junctions.

A note on locality is required at this place. Locality of
the polarization fluctuations (seen as $\tilde{\chi}_i\tilde{\chi}_i$
terms in the general answer) does not generally imply
the locality of the current fluctuations (that would be seen
as $\chi^w_i \chi^b_i$ terms). The reason for this is evident:
(Incomplete) spin current conservation in the node. Let us
give a polarization fluctuation to a connector. It produces not
only a spin current
that goes to the terminal. At the same time, it sends a spin
current (equal in magnitude and opposite in spin direction)
to the node. The node re-distributes this current over all outgoing
channels. Therefore, a local polarization fluctuation
creates the spin current in all transport channels. If spin relaxation
in the node is sufficiently strong, the re-distribution 
ceases ($\chi_N \to 0$) and spin currents are really local.

The third mechanism also does not work for tunnel junctions.
However, it is  one of the two that work for purely 
ballistic connectors
($T_i =1$) that cannot exhibit conductance/polarization fluctuations.
Inspecting (\ref{mech3}), we note that the spin currents
in each connector are no more proportional to local filling
factor drops. Rather, their intensity is set by a single 
global factor $F$ that depends on filling factor drops on all
channels and defines. However,  the fluctuations are local as for LPF
mechanism, with the same reservation. 
The picture behind it is as follows. Let us note that the
node in our model works like a distributor: It receives
electrons coming through transport channels and sends then back.
If it does so with a slight random preference of
the channels with respect to spin, we reproduce the structure
of (\ref{mech2}). Let us call the mechanism {\it local distribution fluctuations}(DF). Similar to LPF, it survives the limit of large $\eta_{SO}$.

The fourth mechanism (\ref{mech4}) does not manifest
itself in charge current fluctuations since it is due to
incompleteness of spin conservation  in the node. As in the case
of LDF, the intensity of current fluctuations in all channels
is set by the global parameter $F$. In distinction from LDF,
the fluctuations are globally orchestrated. It looks like
our node-distributor has some supply of spin of random
direction and sends it uniformly to all channels.
Let us call it {\it global distribution fluctuations}(GDF).
The contribution of this mechanism vanishes upon increasing $\eta_{SO}$.

\section{Tunnel Connectors}
We turn to more concrete and quantitative examples.
Let us assume that all connectors are tunnel
junctions, so that all $T_i \ll 1$. As mentioned,
the fluctuations in this case are due to GPF mechanism.
It is convenient to turn back from transport
channels to terminals, so $i$ now labels the terminals
with conductance $G_i=G_Q\sum_p T^{(i)}_p$. Let us restrict
ourselves to voltage differences $\ll e E_{Th}$. Under
these conditions, we may disregard energy differences
in the diffuson propagators $D_{s,t}$. Integrating
the general answer over energies and taking the limit $T^{(i)}_p \to 0$,
we obtain the following relation for the correlator
of the spin currents:
\begin{equation}
\langle\!\langle I^{\alpha}_i I^{\beta}_j\rangle\!\rangle = \delta_{\alpha\beta}
\frac{I_i I_j} {e^2} \left(\frac{G_Q}{G_\Sigma}\right)^2 \left( \frac{1}{1+\eta_{SO}} - \frac{1}{(1+\eta_{SO})^2}\right)
\end{equation}
Here $I_i$ is the average current to the terminal $i$ determined
by the conductances of the connectors and the voltages applied
in accordance with Ohm's law. We can now explicitly
see that the spin current fluctuation in each connector
is proportional to the current, so that the fluctuating
quantity is the polarization. The full spin current to/from
the system does not fluctuate, although it is generally allowed.
This is specific for tunnel connectors.

It makes sense to compare the spin current fluctuations
with fluctuations of charge currents. Those are given by
\begin{equation}
\langle\!\langle I_i I_j\rangle\!\rangle = 
{I_i I_j} \left(\frac{G_Q}{G_\Sigma}\right)^2 \left( 1 +3 \frac{1}{(1+\eta_{SO})^2}\right)
\end{equation}
As known,\cite{Altshuler,BeenakkerReview}
this fluctuation  reduces by a factor of $4$ upon increasing $\eta_{SO}$.
In RMT language, this is the transition from orthogonal to symplectic
ensemble. 
\begin{figure}
\centerline{\includegraphics[width=12cm]{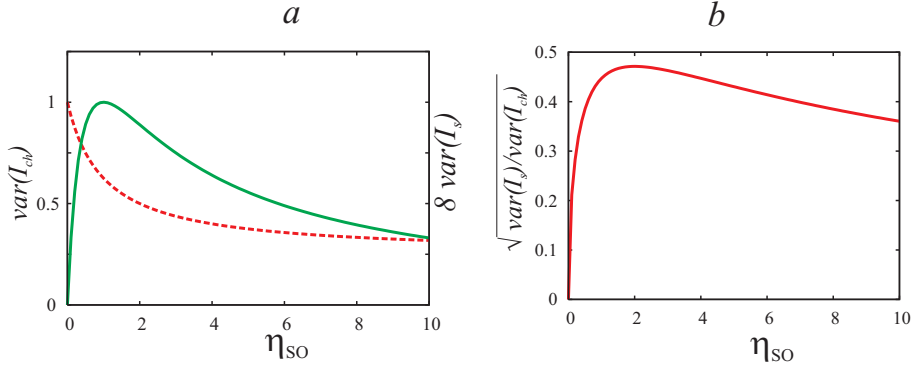}}
\caption{Tunnel connectors a. Variations of charge (dashed line)
and spin currents versus the strength of spin-orbit interaction
$\eta_{SO}$. The variations are normalized by the value
of charge current fluctuations at $\eta_{SO}=0$. Note
factor of $8$ in the scale for spin current fluctuations.
b. Ratio of variations changes slowly in a wide interval
of $\eta_{SO}$.}
\label{tunnel}
\end{figure}
We plot in Fig. 2 the strength of spin and charge fluctuations and
its ratio versus $\eta_{SO}$. Spin fluctuations reach maximum at $\eta_{SO} 
\approx 0.75$. The ratio of the fluctuations reaches maximum at $\eta_{SO} \approx 1.5$ and falls off very slowly upon increasing $\eta_0$. The
polarization of the fluctuating current in maximum is about 50 \%. 
\section{Ballistic Connectors}
Let us turn to the analysis of purely ballistic connectors ($T_i=1$).
As in previous Section, we arrange transport channels into
connectors and label them with $i$. The fluctuations are contributed
by LDF and GDF mechanisms. 
Again we assume $V_i \ll E_{Th}$. 
We integrate the non-equilibrium factor $F$ over energies to obtain
(apart from $e^2$ factor) a strikingly simple expression:
\begin{equation}
\tilde{F} = \sum_i I_i \tilde{V}_i/G_{\Sigma}
\end{equation}
where $\tilde{V}_i$ is the voltage drop at the corresponding connector.
The non-equilibrium factor is thus proportional to the total
energy dissipation in the system.

The factor determines the strength of the spin current fluctuations.
From the general answer (\ref{mech3}) we obtain the contribution of LDF mechanism:
\begin{equation}
\langle\!\langle I^{\alpha}_i I^{\beta}_j\rangle\!\rangle = 
\delta_{\alpha\beta} = \frac{G^2_Q}{e^2 G_\Sigma}\tilde{F} \left( G_i \delta_{ij} - \frac{G_i G_j}{G_\Sigma(1+\eta_{SO})^2}\right) \frac{\eta_{SO}(2+\eta_{SO})}{(1+\eta_{SO})^2}
\end{equation}
We can explicitly see that the fluctuations are local ( with the reservation made above).
It is instructive to give the expression for the charge current
fluctuations:
\begin{equation}
\langle\!\langle I_i I_j\rangle\!\rangle = 
\delta_{\alpha\beta} = \tilde{F} \left( 
G_i \delta_{ij} - \frac{G_i G_j}{G_\Sigma}
\right) 
\left(1+\frac{3}{(1+\eta_{SO})^2}\right)
\end{equation}
The contribution of the GPF mechanism reads:
\begin{equation}
\langle\!\langle I^{\alpha}_i I^{\beta}_j\rangle\!\rangle = 
\delta_{\alpha\beta}  \tilde{F}  \frac{G_i G_j}{8G_\Sigma(1+\eta_{SO})^2}
\frac{\eta^2_{SO}(2+\eta_{SO})}{(1+\eta_{SO})^2}
\end{equation}

In distinction from the tunnel junction system, there is a fluctuation
of the full current to/from the system, $I^{\alpha}_F\equiv\sum_i I^{\alpha}_i$. It simply reads
\begin{equation}
\langle\!\langle I^{\alpha}_i I^{\beta}_j\rangle\!\rangle = 
\frac{G_Q^2}{e^2}\tilde{F}\frac{\eta_{SO}(2+\eta_{SO})}{(1+\eta_{SO})^2}
\end{equation}
where we sum up the contributions of both mechanisms.
\begin{figure}
\centerline{\includegraphics[width=12cm]{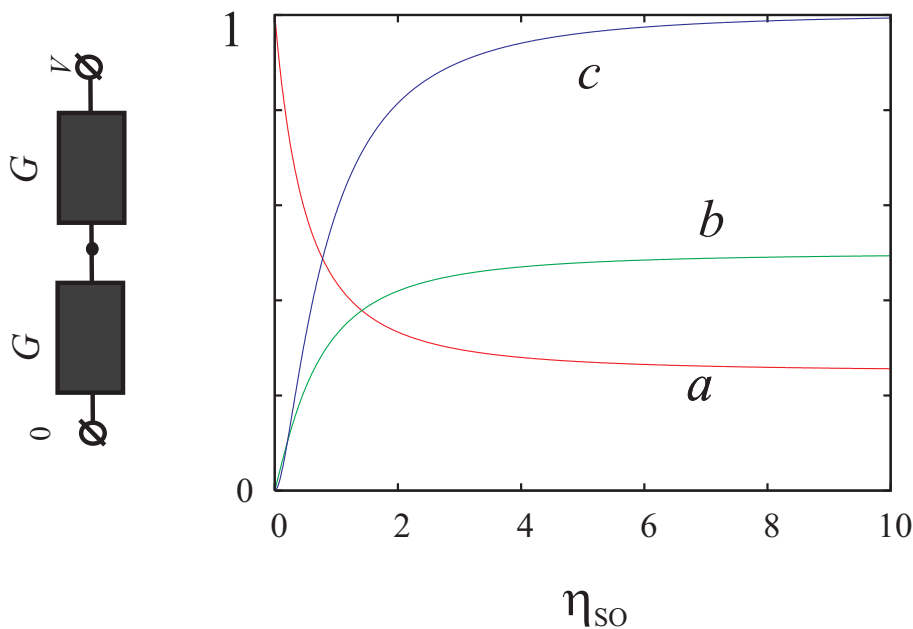}}
\caption{Ballistic connectors. We concentrate
on a simple circuit with identical connectors (on the left).
We plot variances of charge current (a), spin current
in a connector(b), and full spin current to/from the node.
The variances are normalized on the charge current variance
at $\eta_{SO}$.}
\label{ballistic}
\end{figure}
We plot in Fig. 3 the fluctuations of electric
current, full and transport spin current
for a simple system of two connectors with equal conductances.

\section{Conclusions}
To conclude, we have studied the mesoscopic fluctuations
of spin currents for generic model of a single-node, multi-terminal
nanostructure. Spin-orbit interaction is characterized
by a single parameter $\eta_{SO}$ which is the ratio of the
dwell time $\hbar/E_{Th}$ in the node to spin relaxation rate $\tau_{SO}$.
At moderate values of this parameter, the scale of spin current fluctuations
is the same as that of charge current fluctuations.

We have found that the spin current fluctuations are contributed 
by four distinct mechanisms: Fluctuations of polarization or 
distribution, both can be either local or global.
Local mechanisms survive in
the limit of strong spin-orbit interaction.
If all connectors
are of tunnel nature, only GPF works. For purely ballistic connectors,
LDF and GDF contribute. We present simple formulas valid
for tunnel and ballistic connectors and some quantitative results.

\ack The author appreciates useful discussions and communications with G. E. W. Bauer, B. Kramer, J. Ohe and  G. Campagnano.

\end{document}